\documentclass[twoside,11pt]{article}

\usepackage[cp1251]{inputenc}

\usepackage{graphicx}

\usepackage{amssymb}
\usepackage{amsmath}
\usepackage{amsfonts}
\usepackage{wasysym}
\usepackage{revsymb}

\usepackage{titlesec}

\usepackage{cite}

\titleformat*{\section}{\bfseries}
\titleformat*{\subsection}{\bfseries}

\begin{document}           % End of preamble and beginning of text.

\title{\Large 
Uncertainty principle in quantum mechanics with Newton's gravity}
\author{V.E. Kuzmichev, V.V. Kuzmichev\\[0.5cm]
\itshape Bogolyubov Institute for Theoretical Physics,\\
\itshape National Academy of Sciences of Ukraine, Kiev, 03143 Ukraine}

\date{}

\maketitle

\begin{abstract}
A new derivation is given of the known generalized position-momentum uncertainty relation, which takes into account gravity.
The problem of two massive particles, the relative motion of which is described by the Schr\"{o}dinger equation, is considered.
The potential energy is defined as a sum of `standard' non-gravitational term and the second one, which corresponds to gravitational 
attraction of particles as in Newton's theory of gravity. The Green's function method is applied to solve the Schr\"{o}dinger equation.
It is assumed that the solution of the problem in the case, when the gravitational interaction is turned off, is known. Gravity is taken 
into account in linear approximation with respect to the gravitational coupling constant made dimensionless. Dimensional 
coefficients at additional squares of mean-square deviations of position and momentum are written explicitly. The minimum length,
determined as minimal admissible distance between two quantum particles, and the minimum momentum appear to be depending 
on the energy of particles' relative motion. The theory allows one to present
the generalized position-momentum uncertainty relation in a new compact form.
\end{abstract}

PACS numbers: 03.65.-w, 03.65.Ta, 04.60.Bc

\section{Introduction}\label{sec:1}
The Heisenberg uncertainty relation is a fundamental principle of quantum theory. It imposes restriction on accuracy of simultaneous 
measurement of two observables which do not commute \cite{deB,Mes}. At first, gravity was not taken into account. 
The studies conducted by a number of researchers that used different approaches and methods have shown that account of gravity
leads to nontrivial results, which are consistent between themselves (see the reviews of the articles published prior to 1995 in Ref.~\cite{Gar} 
and of the articles published later in Ref.~\cite{Cha,Per}). It was demonstrated, in the context of string theories, that
there must exist so-called minimum physical length, which is of order of Planck's length \cite{Ven,Ama,Kon} (see also Refs.~\cite{Wit,Hos}).
This result was confirmed by other authors. In Ref.~\cite{Mag}, the limit to the measurement of the black-hole-horizon area was discussed, 
bringing one to the conclusion that a minimum observable length emerges naturally from a quantum theory of gravity. The effect of gravity 
was investigated by studying the interaction of the electron with the photon using Newtonian gravity and general relativity, and one arrives
at the generalized uncertainty principle in the form known from  string theories \cite{Adl}. The discussion of consequences of the 
generalized uncertainty relation is also contained in Refs.~\cite{Kem,Cap,Bol,Bam,Das,Ber}. The question of the existence of the so-called
maximum length was raised in connection with the cosmological particle horizon as the maximum measurable length in the Universe \cite{Per}.

In the present paper, the effects of gravity are studied on the example of the quantum problem on relative motion of two massive particles
feeling no external forces. It is supposed that the two-particle quantum system is described by the stationary Schr\"{o}dinger equation
with the potential energy, which in addition to `standard' non-gravitational interaction also takes into account gravitational attraction of particles as 
in Newtonian theory. It is assumed that the solution of the problem in the case, when gravity is neglected, is known and hence the standard 
Heisenberg position-momentum uncertainty relation can be written. The problem is to determine the influence of gravity on the mean-square 
deviations of position and momentum and thus to obtain the generalized position-momentum uncertainty relation. The Green's function method
is applied to solving the Schr\"{o}dinger equation.

The setting of a problem and the basic equations are given in Sect.~2. The particles interaction potential is chosen in the form of two summands:
one is responsible for the non-gravitational part, while the another is described by the Newtonian gravitational potential of the mutual attraction 
of two particles. The method of taking gravity into account is described in Sect.~3. Here, the gravitational coupling constant made dimensionless
is introduced and the normalization condition of the wave function of the system, which takes into consideration Newtonian gravity, 
is written in a linear approximation with respect to this coupling constant. 
In Sect.~4, the generalized position-momentum uncertainty relation is obtained and presented in the form
allowing its direct comparison with the known relation. The restrictions imposed by the minimum length and the minimum momentum 
on the dimensional coefficients at additional squares of mean-square deviations of position and momentum are discussed in Sects.~5 and 6.  
In Sect.~7, the generalized uncertainty relation is 
given in a new compact form in which the gravitational effects are combined into a single summand having a clear physical meaning.
A comment on the Coulomb problem is given here.

\section{Basic concepts and equations}\label{sec:2}
We consider the quantum system of two non-relativistic particles with the masses $m_{1}$ and $m_{2}$ and the respective positions 
$\mathbf{r}_{1}$ and $\mathbf{r}_{2}$, which interact between themselves. We also assume that the system feels no external forces.
As is well-known \cite{Mes}, the transition to the center-of-mass system allows one to split up the problem into two: that of 
a free particle with the mass $M = m_{1} + m_{2}$, and that of a particle with the reduced mass $m = m_{1} m_{2}/ M$ in some static
potential. We take the interaction potential energy in the form of a sum $V(\mathbf{r}) + U(r)$, where $V(\mathbf{r})$ is the non-gravitational 
potential energy, which depends only upon the relative position $\mathbf{r} = \mathbf{r}_{1} - \mathbf{r}_{2}$, and $U(r)$ is the potential 
energy of gravitational attraction of particles, 
\begin{equation}\label{1}
U(r) = - G \frac{m_{1} m_{2}}{r} = - G \frac{m M}{r} = m \Phi(r),
\end{equation}
where $r = |\mathbf{r}|$, $\Phi(r) = - G M/r$ is the gravitational potential created by the mass $M$ acting on the particle of mass $m$ 
at a distance $r$ from the force-center,  and $G$ is Newton's gravitational constant.

The Schr\"{o}dinger equation for relative motion of the particles has the form
\begin{equation}\label{2}
\left(H_{0} - E \right) |\psi \rangle = - U |\psi \rangle,
\end{equation}
where $\langle \mathbf{r} |\psi \rangle = \psi (\mathbf{r})$ is the wave function (in the coordinate representation), $E$ is the energy of 
relative motion, and
\begin{equation}\label{3}
H_{0} = \frac{\mathbf{p}^{2}}{2 m} + V(\mathbf{r})
\end{equation}
is the Hamiltonian of a particle of mass $m$ and momentum $\mathbf{p} = (p_{x}, p_{y}, p_{z})$, which is canonically conjugate to 
position $\mathbf{r} = (x,y,z)$. This Hamiltonian does not take into account the gravitational impact of the force-center on this particle.
As it follows from Eq.~(\ref{1})-(\ref{3}), in quantum mechanics the mass $m$ no longer cancels, in contrast to classical theory of gravity,
where the mass of a body does not appear in the equation describing its motion in external gravitational field \footnote{This difference 
between classical and quantum theories of gravity is well-known (see, e.g., \cite{Sac94}).}. In quantum theory, this fact can be used for 
transition from the dimensional constant of gravitational interaction $G$ to the dimensionless coupling constant
\begin{equation}\label{4}
g \equiv G \frac{m_{1} m_{2}}{\hbar c} = G \frac{m M}{\hbar c},
\end{equation}
so that Eq.~(\ref{1}) takes the form
\begin{equation}\label{5}
U(r) = - g \frac{\hbar c}{r}.
\end{equation}
The condition of ``smallness'' of the constant $G$ reduces to a mathematically correct condition: $g \ll 1$, which is equivalent to
inequalities,
\begin{equation}\label{6}
\frac{l_{pl}}{\lambdabar_{m}} \frac{l_{pl}}{\lambdabar_{M}} \ll 1, \quad \mbox{or} \quad \frac{m}{m_{pl}} \frac{M}{m_{pl}} \ll 1,
\end{equation}
where $l_{pl} = \sqrt{\frac{G \hbar}{c^{3}}}$ is Planck's length, $m_{pl} = \frac{\hbar}{l_{pl} c}$ is Planck's mass, 
$\lambdabar_{m} = \frac{\hbar}{m c}$ is Compton's wavelength of a particle with the mass $m$ (the same is for $M$).
This restriction means that the masses $m$ and/or $M$ (and correspondingly $m_{1}$ and $m_{2}$) must be smaller than
Planck's mass.

\section{Account of gravity}\label{sec:3}
We suppose that the solution of Eq.~(\ref{2}) without right-hand side 
\begin{equation}\label{7}
\left(H_{0} - E \right) |\varphi \rangle = 0
\end{equation}
is known, as well as the Heisenberg uncertainty relation, for example, for a particle that moves along the $x$-direction,
\begin{equation}\label{8}
\delta x \delta p_{x} \geq \frac{\hbar}{2},
\end{equation}
where $\delta x = \sqrt{\bar{x^{2}} - \bar{x}^{2}}$ and $\delta p_{x} = \sqrt{\bar{p_{x}^{2}} - \bar{p_{x}}^{2}}$ are the root-mean-square 
deviations of position and momentum respectively, and the over-bar denotes averaging over the state $|\varphi \rangle$. This relation
does not account for the gravitation attraction of particles.

Let us calculate the gravitational correction to the relation (\ref{8}), using the solution of Eq.~(\ref{2}) for this purpose.
The general solution of Eq.~(\ref{2}) can be written as
\begin{equation}\label{9}
|\psi \rangle = |\varphi \rangle + |\chi \rangle,
\end{equation}
where $|\varphi \rangle$ is the general solution of Eq.~(\ref{7}), and the function $|\chi \rangle$ has a form
\begin{equation}\label{10}
|\chi \rangle = G_{E} U |\psi \rangle \quad \mbox{and} \quad \langle \chi | = \langle \psi | U G_{E}^{\dagger},
\end{equation}
where
\begin{equation}\label{11}
G_{E} = \sum_{n} \frac{| \varphi_{n} \rangle \langle \varphi_{n} |}{E + i \delta - E_{n}},
\end{equation}
is the Green's function for outgoing waves, $\delta$ is a small positive addition, and the Hermitian-conjugate function
$G_{E}^{\dagger}$ describes incoming waves. Here, the wave function $| \varphi_{n} \rangle$ and the corresponding
energy $E_{n}$ are the solutions of the equation
\begin{equation}\label{12}
\left(H_{0} - E_{n} \right) |\varphi_{n} \rangle = 0
\end{equation}
for eigenvalues and eigenvectors. In Eq.~(\ref{11}), the summation over discrete states and integration over continuum states of
the two-particle system in which gravity is not taken into account is implied.

We assume that the wave functions $|\varphi_{n} \rangle$ form a complete orthonormalized set: $\sum_{n} | \varphi_{n} \rangle \langle \varphi_{n} | = 1$,
$\langle \varphi_{n} | \varphi_{n'} \rangle = \delta_{nn'}$. Using these conditions, one can calculate the norm of the state $| \varphi \rangle$.
From the obvious expansion
\begin{equation}\label{13}
| \varphi \rangle = \sum_{n} | \varphi_{n} \rangle \langle \varphi_{n} | \varphi \rangle,
\end{equation}
it follows that
\begin{equation}\label{14}
\langle \varphi | \varphi \rangle = \sum_{n} | \langle \varphi_{n} | \varphi \rangle |^{2} = 1,
\end{equation}
since $| \langle \varphi_{n} | \varphi \rangle |^{2}$ is the probability to find the system with the Hamiltonian (\ref{3}) in the state described by the wave function 
$|\varphi_{n} \rangle$.

Using Eqs.~(\ref{9}), (\ref{10}), and (\ref{14}), we obtain an exact expression for the norm $\langle \psi | \psi \rangle$,
\begin{equation}\label{15}
\langle \psi | \psi \rangle = 1 + \langle \varphi | G_{E} U | \psi \rangle + \langle \psi | U G_{E}^{\dagger} | \varphi \rangle + 
\langle \psi | (G_{E} U)^{\dagger} G_{E} U | \psi \rangle.
\end{equation}
By substituting the wave function (\ref{9}) into Eq.~(\ref{15}), we obtain the series in the form of expansion in powers of a coupling constant $g$. 
Keeping only the linear term in this expansion, with an accuracy up to $\sim O(g^{2})$ we find
\begin{equation}\label{16}
\langle \psi | \psi \rangle = 1 + \overline{G_{E} U} + \overline{U G_{E}^{\dagger}},
\end{equation}
where the over-bar denotes averaging over the state $|\varphi \rangle$ as above. We rewrite the nontrivial part of Eq.~(\ref{16}) in an explicit form,
\begin{equation}\label{17}
\begin{split}
\overline{G_{E} U} + \overline{U G_{E}^{\dagger}} & = \sum_{n}\!' \frac{1}{E - E_{n}} \left[\langle \varphi | \varphi_{n} \rangle \langle \varphi_{n} | U | \varphi \rangle
+ \langle \varphi | U | \varphi_{n} \rangle \langle \varphi_{n} | \varphi \rangle \right] \\ 
& - i \pi \sum_{n} \delta(E - E_{n}) \left[\langle \varphi | \varphi_{n} \rangle \langle \varphi_{n} | U | \varphi \rangle
- \langle \varphi | U | \varphi_{n} \rangle \langle \varphi_{n} | \varphi \rangle \right],
\end{split}
\end{equation}
where a dash denotes that the term with $E = E_{n}$ is absent in the summation. Since the left-hand side of Eq.~(\ref{17}) is a real number,
then in the right-hand side the contributions from incoming and outgoing waves should cancel mutually. As a result, we obtain the expression
\begin{equation}\label{18}
\overline{G_{E} U} + \overline{U G_{E}^{\dagger}} = 2 \sum_{n}\!' \frac{1}{E - E_{n}} \langle \varphi | \varphi_{n} \rangle \langle \varphi_{n} | U | \varphi \rangle
\equiv 2 \overline{G'_{E} U},
\end{equation}
which does not contain a singularity at the point $E = E_{n}$.

As $U < 0$ for all $r$, the norm (\ref{16}) is less than unity: $\langle \psi | \psi \rangle < 1$.

\section{Generalized uncertainty relation}\label{sec:4}
Let us calculate the root-mean-square deviations of position $x$, $\Delta x = \sqrt{\langle x^{2} \rangle - \langle x \rangle^{2}}$, and momentum 
$p_{x}$, $\Delta p_{x} = \sqrt{\langle p_{x}^{2} \rangle - \langle p_{x} \rangle^{2}}$, where the brackets denote averaging over the state $| \psi \rangle$
(\ref{9}). For the mean value of some Hermitian operator $A$, with an accuracy up to terms $\sim O(g^{2})$, we have
\begin{equation}\label{19}
\langle A \rangle \equiv \frac{\langle \psi | A | \psi \rangle}{\langle \psi | \psi \rangle} = \overline{A} + \overline{A G_{E} U} + \overline{U G_{E}^{\dagger} A}
- \overline{A} \left(\overline{G_{E} U} + \overline{U G_{E}^{\dagger}} \right).
\end{equation}
By setting $A$ equal to $x$, $x^{2}$, $p_{x}$, and $p_{x}^{2}$ sequentially, we get
\begin{equation}\label{20}
(\Delta x)^{2} = (\delta x)^{2} + \beta, \quad (\Delta p_{x})^{2} = (\delta p_{x})^{2} + \alpha,
\end{equation}
where we denote
\begin{equation}\label{21}
\beta = - (\delta x)^{2} \left(\overline{G_{E} U} + \overline{U G_{E}^{\dagger}} \right) + R_{x}, \quad
\alpha = - (\delta p_{x})^{2} \left(\overline{G_{E} U} + \overline{U G_{E}^{\dagger}} \right) + R_{p_{x}},
\end{equation}
and
\begin{equation}\label{22}
R_{x} \equiv \overline{x}^{2} \left(\overline{G_{E} U} + \overline{U G_{E}^{\dagger}} \right) - 
2 \overline{x} \left(\overline{x G_{E} U} + \overline{U G_{E}^{\dagger} x} \right) +
\overline{x^{2} G_{E} U} + \overline{U G_{E}^{\dagger} x^{2}}
\end{equation}
(the same expression is for $R_{p_{x}}$).

Neglecting the term $\alpha \beta \sim O(g^{2})$, one can write
\begin{equation}\label{23}
(\Delta x)^{2} (\Delta p_{x})^{2} = (\delta x)^{2} (\delta p_{x})^{2} + \alpha (\delta x)^{2} + \beta (\delta p_{x})^{2}.
\end{equation}
Bearing in mind the equations (\ref{20}) and the uncertainty relation (\ref{8}), we obtain the following inequality, with the accuracy mentioned above,
\begin{equation}\label{24}
(\Delta x)^{2} (\Delta p_{x})^{2} \gtrsim \frac{\hbar^{2}}{4} + \alpha (\Delta x)^{2} + \beta (\Delta p_{x})^{2}.
\end{equation}
After taking the root, we arrive at the required position-momentum uncertainty relation, which takes into account the gravitational corrections
in linear approximation with respect to the coupling constant $g$,
\begin{equation}\label{25}
\Delta x \frac{\Delta p_{x}}{\hbar} \gtrsim \frac{1}{2} + \frac{\alpha}{\hbar^{2}} (\Delta x)^{2} + \beta \left( \frac{\Delta p_{x}}{\hbar}\right)^{2}.
\end{equation}
The constant $\frac{\alpha}{\hbar^{2}}$ has dimension of inverse length squared, and $\beta$ is measured in units of length squared.
The inequality (\ref{25}) has the same form as the generalized uncertainty relation proposed previously by many authors 
(see introduction to this paper). In our approach, the coefficients $\alpha$ and $\beta$ can be calculated by the formulas (\ref{21}) and (\ref{22}).
We shall consider these coefficients in detail.

Keeping in mind the mean value theorem, let us suppose that there exist $x_{c}$ and $(x^{2})_{c}$ such that
an exact expression for the integrals in Eq.~(\ref{22}) can be written as
\begin{equation}\label{26}
\overline{x G_{E} U} = x_{c} \overline{G_{E} U}, \quad \overline{x^{2} G_{E} U} = (x^{2})_{c} \overline{G_{E} U}.
\end{equation}
The same is for the Hermitian conjugates. Then
\begin{equation}\label{27}
R_{x} = 2 \overline{G'_{E} U} \left[\overline{x}^{2} - 2 \overline{x} x_{c} + (x^{2})_{c} \right],
\end{equation}
where denotations are as in Eq.~(\ref{18}). Assuming that $(x^{2})_{c} \approx (x_{c})^{2}$, we have
\begin{equation}\label{28}
R_{x} \approx 2 \overline{G'_{E} U} (\delta x)^{2}_{c},
\end{equation}
where $(\delta x)^{2}_{c} = \left(x_{c} - \overline{x} \right)^{2}$. The same expressions can be written for $R_{p_{x}}$ as well,
\begin{equation}\label{29}
R_{p_{x}} \approx 2 \overline{G'_{E} U} (\delta p_{x})^{2}_{c},
\end{equation}
where $(\delta p_{x})^{2}_{c} = \left(p_{c} - \overline{p_{x}} \right)^{2}$. Since the deviations $(\delta x)_{c}$ and $(\delta p_{x})_{c}$
are not directly related to each other through the Heisenberg uncertainty principle, they can become arbitrarily small simultaneously.
Then, in a good approximation, we get
\begin{equation}\label{30}
\beta = 2 g (\delta x)^{2} \overline{G'_{E} \frac{\hbar c}{r}}, \quad \alpha =  2 g (\delta p_{x})^{2} \overline{G'_{E} \frac{\hbar c}{r}}.
\end{equation}

\section{Minimum length}\label{sec:5}
Let us consider the case, when the summand containing $\alpha$ is much smaller than the one containing $\beta$.
From the uncertainty condition (\ref{8}) and an obvious condition 
\begin{equation}\label{31}
\delta p_{x} \lesssim p_{x} = \frac{\hbar}{\lambdabar_{m}},
\end{equation}
it follows that
\begin{equation}\label{32}
\delta x \geq \frac{\hbar}{2 \delta p_{x}} \gtrsim \frac{\hbar}{2 p_{x}} = \frac{\lambdabar_{m}}{2}.
\end{equation}
Using an explicit form of the coupling constant $g$ (\ref{4}), we obtain
\begin{equation}\label{33}
\beta \gtrsim l_{pl}^{2} \frac{\beta'}{2},
\end{equation}
where
\begin{equation}\label{34}
\beta' = \frac{\lambdabar_{m}}{\lambdabar_{M}} \overline{G'_{E} \frac{\hbar c}{r}}
\end{equation}
is a dimensionless coefficient which depends on the free parameters of the two-particle system under consideration. 
Then, using the equations (\ref{33}), the uncertainty relation (\ref{25}) can be reduced to the form
\begin{equation}\label{35}
\Delta x \gtrsim \frac{\hbar}{2 \Delta p_{x}} + l_{pl}^{2} \frac{\beta'}{2} \frac{\Delta p_{x}}{\hbar}.
\end{equation}
From this inequality, it follows that for the system of two massive particle under consideration, there exists the minimal value of deviation
$\Delta x$, which is determined by the coefficient $\beta'$,
\begin{equation}\label{36}
(\Delta x)_{min} = l_{pl} \sqrt{\beta'} \equiv l_{min},
\end{equation}
at the momentum deviation $\Delta p_{x} = \frac{\hbar}{l_{pl} \sqrt{\beta'}}$. The quantity $\frac{\hbar c}{l_{pl}^{2} \beta'}$ has the dimension of energy 
divided by length and it is associated with the string tension in string theory. The distances smaller than the minimum length  $l_{min}$ cannot be resolved.
According to Eq.~(\ref{34}), the quantity $\beta'$ depends on the energy $E$ of relative motion of the particles with masses $m$ and $M$.
If there exist the values $E \sim E_{0}$ at which  $\beta' \sim 1$, then the minimum length $l_{min} \sim l_{pl}$.
In the general case, $\sqrt{\beta'}  = \frac{l_{min}}{l_{pl}}$, where $l_{min}$ depends on energy as
\begin{equation*}
l_{min} \sim l_{pl} \left[\frac{\hbar c}{E} \frac{M}{m} \overline{\frac{1}{r}}\right]^{1/2},
\end{equation*} 
in a rough approximation.

The relation (\ref{35}) has two limits (cf. Ref.~\cite{Bol}). The quantum-mechanical limit is achieved when 
\begin{equation}\label{37}
l_{pl}^{2} \beta' \left(\frac{\Delta p_{x}}{\hbar}\right)^{2} \ll 1, \quad \mbox{or} \quad \frac{\Delta p_{x}}{\hbar} \ll \frac{1}{l_{min}}.
\end{equation}
The standard Heisenberg relation is recovered in the case $\Delta x \gg l_{pl}$. 

The quantum-gravity limit is reached when
\begin{equation}\label{38}
l_{pl}^{2} \beta' \left(\frac{\Delta p_{x}}{\hbar}\right)^{2} \sim 1, \quad \mbox{or} \quad \frac{\Delta p_{x}}{\hbar} \sim \frac{1}{l_{min}}.
\end{equation}
This limit means that, according to Eq.~(\ref{35}), 
\begin{equation}\label{39}
\frac{\Delta p_{x}}{\hbar} \sim \frac{\Delta x}{l_{pl}^{2} \beta'},
\end{equation} 
which transforms into the relation (\ref{38}) for $\Delta x \sim l_{pl}$.

\section{Minimum momentum}\label{sec:6}
When the summand containing $\beta$ is much smaller than the one with $\alpha$, the uncertainty relation (\ref{25}) takes the form
\begin{equation}\label{40}
\Delta x \frac{\Delta p_{x}}{\hbar} \gtrsim \frac{1}{2} \left[1 + \frac{\alpha'}{l_{pl}^{2}} (\Delta x)^{2} \right],
\end{equation}
where it is denoted $\alpha' = \frac{2 \alpha l_{pl}^{2}}{\hbar^{2}}$. One obtains the quantum-mechanical limit of this relation in the case
\begin{equation}\label{41}
\frac{\alpha'}{l_{pl}^{2}} (\Delta x)^{2} \ll 1, \quad \mbox{or} \quad \Delta x \ll \frac{l_{pl}}{\sqrt{\alpha'}}.
\end{equation} 
One gets the quantum-gravity limit when
\begin{equation}\label{42}
\frac{\alpha'}{l_{pl}^{2}} (\Delta x)^{2} \sim 1, \quad \mbox{or} \quad \Delta x \sim \frac{l_{pl}}{\sqrt{\alpha'}}.
\end{equation} 
For $\delta p_{x} \approx p_{x} = \frac{\hbar}{\lambdabar_{m}}$, it follows that
\begin{equation}\label{43}
\sqrt{\alpha'}  \approx 2 \left(\frac{l_{pl}}{\lambdabar_{m}}\right)^{2} \sqrt{\beta'} = 2 \left(\frac{m}{m_{pl}}\right)^{2} \sqrt{\beta'}.
\end{equation} 
For masses $m \ll m_{pl}$ and a finite value of $\sqrt{\beta'}$, one finds that $\sqrt{\alpha'} \ll 1$.
The equation (\ref{42}) can be rewritten as
\begin{equation}\label{44}
\Delta x \sim \frac{1}{2} \left(\frac{m_{pl}}{m}\right)^{2} \frac{l_{pl}^{2}}{l_{min}}.
\end{equation} 
The quantum gravitational effects will manifest themselves at very large distances \cite{Kuz19} (cf. Ref.~\cite{Bol}) for particles with masses 
$m \ll m_{pl}$.

Let us make a few estimates. Taking $l_{min} \sim l_{pl}$, one finds that $\Delta x \sim 10^{5}$ cm for $m c^{2} \sim 1$ GeV.
On the other hand, under the assumption that the size of fluctuations are of order the observable part of the Universe
(the maximum measurable length in the Universe) $\Delta x \sim 10^{28}$ cm, one obtains that masses of particles $m c^{2} \sim 10^{-3}$ eV.
Current upper limits on neutrino mass \cite{Mer} agrees with this estimate.

From the uncertainty relation (\ref{40}) in a lower limit, it follows that
\begin{equation}\label{45}
\frac{\alpha' \hbar^{2}}{l_{pl}^{2} (\Delta p_{x})^{2}} \lesssim 1, \quad \mbox{or} \quad \frac{\Delta p_{x}}{\hbar} \gtrsim \frac{\sqrt{\alpha'}}{l_{pl}}.
\end{equation} 
Thus, there exists a minimum momentum $(\Delta p_{x})_{min} = m_{pl} c \sqrt{\alpha'}$.

\section{Discussion}\label{sec:7}
In this paper, it is shown how, using the Schr\"{o}dinger equation for two massive particles with Newton's gravity, one can obtain the
generalized uncertainty relation in the form which became known since the second half of 1980s (see Eqs.~(\ref{25}) and (\ref{35})).
In linear approximation with respect to the dimensionless coupling constant $g$ (\ref{4}) of gravitational interaction, 
according to Eqs.~(\ref{33}) and (\ref{34}), the coefficients $\alpha$ and $\beta$ are expressed in the form of quadratures containing 
the Green's function and the gravitational potential, and they depend on the energy of relative motion of the particles.
The minimum length of the order of the Planck length can be achieved for some values of the energy $E$, which is a free parameter of the theory.
In the framework under consideration, the uncertainty relation can be cast into a new form. Starting from Eq.~(\ref{23}), using 
the coefficient definitions (\ref{21}) and the Heisenberg uncertainty relation (\ref{8}), we arrive at the inequality
\begin{equation}\label{46}
\Delta x \frac{\Delta p_{x}}{\hbar} \gtrsim \frac{1}{2} - \frac{1}{2} \left(\overline{G_{E} U} + \overline{U G_{E}^{\dagger}} \right) =
\frac{1}{2} + g \overline{G'_{E} \frac{\hbar c}{r}}.
\end{equation}
This relation, in turn, can be reduced to a compact form by taking into account the normalization condition (\ref{16}) in linear approximation 
with respect to the coupling constant $g$,
\begin{equation}\label{47}
\Delta x \Delta p_{x} \gtrsim \frac{\hbar}{2} \frac{1}{\langle \psi | \psi \rangle}.
\end{equation}

The potential energy of gravitational attraction of two massive particles (\ref{1}) is similar to the potential energy of two oppositely charged
point particles interacting  according to the Coulomb law. 

Applying the method given above to the Coulomb problem in non-relativistic approximation, we demonstrate that an addition to $\frac{1}{2}$
in Eq.~(\ref{46}) is negligibly small and does not affect the Heisenberg uncertainty relation. Bearing in mind Eq.~(\ref{18}), we denote
this addition as
\begin{equation}\label{48}
\gamma = - \overline{G'_{E} U}.
\end{equation}
Then the coefficient $\beta$ (\ref{33}) takes the form
\begin{equation}\label{49}
\beta \gtrsim \frac{\lambdabar_{m}^{2}}{2}  \gamma.
\end{equation}
Therefore, in order to analyze the contribution from the long-range part of interaction between particles, it is sufficient to estimate the contribution
from a dimensionless coefficient $\gamma$.

As a specific example, we consider the finite motion of an electron with the charge $-e$ in the field of an atomic nucleus with the charge $Ze$
(a hydrogen-like atom). For the light nuclei with $Z \lesssim 10$, the dimensionless coupling constant of the Coulomb interaction is small:
$g = \frac{Z e^{2}}{\hbar c} = \frac{Z}{137} \ll 1$.

As is well-known (see, e.g. Ref.~\cite{Flu}) the energy of electron-nucleus interaction can be written as
\begin{equation}\label{50}
\frac{Z e^{2}}{R^{3}} \left(r^{2} - 3 R^{2} \right) \Theta (R - r) - \frac{Z e^{2}}{r} \Theta (r - R),
\end{equation}
where $\Theta (x) = 1$ at $x > 0$, and $\Theta (x) = 0$ at $x < 0$. The first term describes the potential energy inside the nucleus
with the radius $R$, in which an electric charge is distributed homogeneously over its volume. We identify this term with $V(\mathbf{r})$ in
Eq.~(\ref{3}). The second term corresponds to an additional Coulomb attraction at distances exceeding the size of a nucleus. We set it equal to
$U(r)$.

For the specific problem under consideration, using Eq.~(\ref{18}), (\ref{48}), and (\ref{50}), one can write
\begin{equation}\label{51}
\gamma = Z e^{2} \sum_{n}\!' \frac{\langle \varphi | \varphi_{n} \rangle}{E - E_{n}}  \langle \varphi_{n} | \frac{\Theta (r - R)}{r} | \varphi \rangle.
\end{equation}
The wave functions $\varphi (\mathbf{r})$ and $\varphi_{n}(\mathbf{r})$ are defined over all space of $\mathbf{r} = (x,y,z)$. Since the radius
of a nucleus $R \simeq 2$ fm is much smaller than the Bohr radius of an electron $a = \frac{\hbar^{2}}{m e} = 5.3 \times 10^{4}$ fm,
the electron remains mostly outside the nucleus. In the region $r > R$, the function $\varphi$ has the Coulomb asymptotics, while the 
function  $\varphi_{n}$ decreases outside the nucleus in a standard way. We restrict ourselves to consideration of the $1s$ states. Then, 
\begin{equation}\label{52}
\varphi(\mathbf{r}) = \sqrt{\frac{\varkappa^{3}}{\pi}} e^{- \varkappa r}, \quad
\varphi_{n}(\mathbf{r}) = \sqrt{\frac{\varkappa_{n}}{2 \pi}} \frac{e^{- \varkappa_{n} r}}{r},
\end{equation}
where $\varkappa = \frac{1}{\hbar} \sqrt{2 m |E|}$ and $\varkappa_{n} = \frac{1}{\hbar} \sqrt{2 m |E_{n}|}$.

For the Coulomb problem, according to Eq.~(\ref{9}), the wave function $\psi$ describes exactly, while the wave function $\varphi$
describes approximately the bound states of an electron and a nucleus.

The spectrum of the states of $E$ is close to the spectrum of the states of a hydrogen-like atom. Then, the following estimation is valid:
$\varkappa \approx 1.9 \frac{Z}{\nu} \times 10^{-5}$ fm$^{-1}$, where $\nu = 1,2,3 \ldots$ is the number of a state.
The nuclear potential in Eq.~(\ref{50}) is cut off at $r = R < \sqrt{3} R$. Therefore, the spectrum of the states of $E_{n}$ can be
approximated by the spectrum of the states of a particle in a square well with a radius $R$. In this case, for the values of
$n$ not large enough, we have the estimation: $\varkappa_{n} \approx \frac{\pi}{2} \frac{n}{R}$.

Substituting the wave functions (\ref{52}) into Eq.~(\ref{51}), we obtain
\begin{equation}\label{53}
\gamma \approx 16 Z \frac{R}{a} \lambda^{3} \sum_{n}\!' 
\frac{\lambda_{n} e^{-(\lambda_{n} + \lambda)}}{(\lambda_{n}^{2} - \lambda^{2}) (\lambda_{n} + \lambda)^{3}},
\end{equation}
where we introduce the dimensionless parameters $\lambda = \varkappa R = 3.8 \frac{Z}{\nu} \times 10^{-5}$,
$\lambda_{n} = \varkappa_{n} R \approx \frac{\pi}{2} n$. For $Z \leq 10$, we have $\lambda \ll \lambda_{n}$ and
the following formula for $\gamma$ can be written,
\begin{equation}\label{54}
\gamma \approx 16 Z \frac{R}{a} \lambda^{3} \left(\frac{2}{\pi}\right)^{4} \sum_{n} \frac{e^{- \frac{\pi}{2} n}}{n^{4}}.
\end{equation}
The summand with $n = 1$ gives the main contribution to the sum in Eq.~(\ref{54}), while the summands with $n > 1$ are
exponentially small. After substituting the numerical values for the parameters, we get the following expression for the 
lowest state with $\nu = 1$ which gives the main contribution to $\gamma$,
\begin{equation}\label{55}
\gamma \approx 1.1 Z^{4} \times 10^{-18}.
\end{equation}
For $Z \leq 10$, it follows an upper bound
\begin{equation}\label{56}
\gamma < 10^{-14}.
\end{equation}
Such an addition may be neglected in comparison with $\frac{1}{2}$ in Eq.~(\ref{46}). As a consequence, the standard Heisenberg uncertainty relation
remains valid. It is not violated by the long-range action of the Coulomb attraction between an electron and a nucleus.

Let us note that Eq.~(\ref{2}) with the Hamiltonian (\ref{3}), where $V(\mathbf{r}) = 0$, has an exact solution (a hydrogen-like atom or electron-atomic 
nucleus scattering). The wave function $\psi$ can be normalized to unity (for scattering problem, $\psi$  is represented by the wave packet
formed by superposition of unbound states \cite{Mes}). Then, from Eq.~(\ref{47}), it follows immediately the Heisenberg uncertainty relation
in its standard form.

In conclusion, we mention that the main difference between the gravitational problem and the Coulomb problem is that in the first case the expected new 
quantum effects are to be observed, first of all, at Planck scale, whereas in the second case the relevant scales are atomic or nuclear scales, where
such effects do not manifest themselves on the fluctuations of observed quantities.

\vskip3mm \textit{The present work was partially supported by The National Academy of
Sciences of Ukraine (projects Nos.~0117U000237 and  0116U003191).}

\end{document}